\documentclass[pdflatex,sn-basic]{sn-jnl}

\usepackage{graphicx}%
\usepackage{multirow}%
\usepackage{amsmath,amssymb,amsfonts}%
\usepackage{amsthm}%
\usepackage{mathrsfs}%
\usepackage[title]{appendix}%
\usepackage{xcolor}%
\usepackage{textcomp}%
\usepackage{manyfoot}%
\usepackage{booktabs}%
\usepackage{algorithm}%
\usepackage{algorithmicx}%
\usepackage{algpseudocode}%
\usepackage{listings}%

\theoremstyle{thmstyleone}%
\theoremstyle{thmstyletwo}%

\theoremstyle{thmstylethree}%

\begin{document}

\title[Article Title]{Outstanding Questions in Giant Planet Theory}


\author*[1]{\fnm{Ravit} \sur{Helled}}\email{ravit.helled@uzh.ch}

\affil*[1]{\orgdiv{Department of Astrophysics}, \orgname{University of Zurich}, \orgaddress{\street{Winterthurerstr. 190}, \city{Zurich}, \postcode{8057}, \state{Zurich}, \country{Switzerland}}}


\abstract{Giant planets have key role in shaping planetary systems. Their composition reveals information on the conditions at which planets form, and their interiors serve as natural laboratories to explore the behavior of materials at extreme conditions. They can also host large regular moons that can be habitable. In addition, outside the solar system, giant exoplanets remain the ideal planets for detection and characterization. 
However, despite decades of investigations, and much progress on both the theoretical and observational fonts, several key open questions remain unanswered. In this short review, I highlight a few open questions in the field with the hope that they can be addressed with future research and observational data.}




\maketitle

\section{Introduction}\label{sec1}

 Giant planets serve as benchmarks for studying planetary formation, internal structure, and long-term thermal evolution, within the Solar System and beyond. 
In recent years, substantial progress has been achieved through space missions within the Solar System,  improved equations of state calculations, high-pressure experiments, and increasingly advanced planet formation and evolution simulations. At the same time, the rapid expansion of exoplanet discoveries from space and the ground, with highly improved accuracy, has revealed a wide range of giant planet masses, radii, compositions. 
This large variety challenges classical theoretical frameworks of planet formation and evolution. Despite many advances, several fundamental questions remain unresolved. These include, to mention just a few, the dominant formation pathways of giant planets, the size distribution of solids that are accreted, the path to form giant planets around M-stars, and the link between atmospheric and bulk composition. 
This paper highlights some (but clearly not all) of these critical open questions in giant planet theory. 
\section{Open questions}
\subsection{Can the core accretion model lead to the formation of massive giant planets?}\label{sec2}
The leading paradigm for giant planet formation is the  core accretion (CA) model. It is characterized by a bottom-up formation process in which a solid/heavy-element core grows to a critical mass, typically 5–15 M$_{\oplus}$, before undergoing runaway gas accretion  (\cite{LambrechtsJohansen2012, Ikoma2025}, although see  \cite{Helled2023} for a different perspective). However, as observations of exoplanetary systems reveal "super-Jupiters" with masses exceeding 5–10 M$_J$, the applicability of CA to these massive companions remains a subject of intense debate \citep{2024RvMG...90...55M, JohnstonPanicLiu2023}. 
In particular, forming planets with masses larger than $\sim$ 1  M$_J$
  requires a vast reservoir of gas; whether disk-limited accretion can supply such high masses is questionable, particularly as the planet begins to open a gap in the disk, which inherently restricts further mass gain \citep{DAngeloWeidenschillingLissauerBodenheimer2020}. Super-massive gaseous planets could be a result of collisions between lower-mass giant planets but this mechanism is expected to be rare and requires specific conditions \citep{2021MNRAS.501.1621L}.  
Finally, the observed correlation between host star metallicity and giant planet occurrence is often thought to be a hallmark of the CA model.  However, this trend appears to vanish for companions more massive than approximately 4-10 M$_J$, suggesting a potential transition in formation pathways \citep{Santos2017}. It is currently unknown whether these massive objects represent the high-mass tail of the CA process or if they are formed via Disk Instability (DI), which is expected to be less sensitive to stellar metallicity \citep{Boley2009}.


\subsection{What are the expected compositions of giant planets in the disk instability model?}\label{sec2}
In the disk instability (DI) model, giant planets form via a gravitational instability in the protoplanetary disk. Naively, it is often thought that the DI model leads to planets with stellar composition. In other words, that the composition of the planetary fragment should reflect the bulk composition of the parent nebula. However, given the diversity in formation environments and the various chemical and physical processes that take place during the planetary formation, the composition  could differ significantly in terms of metallicity, heavy-element  distribution, and possibly also isotopic ratios \citep{Helledetal2014}. 
For example, once a DI fragment forms, it can continue to accrete solids from the surrounding disk, potentially leading to moderate metal enrichment after the initial collapse. 
Disk-clump interactions can also change the composition: solids can concentrate in spiral arms before fragmentation occurs, leading to fragments that are born with a slight metal surplus compared to the host star \citep{BoleyHelledPayne2011}. It should also be noted that DI planets can have cores through the settling and sedimentation of heavy elements towards the planetary center \citep{NayakshinHelledBoley2014}.

The expected volatile enrichment and C/O ratios in DI planets remain unknown \citep[e.g.,][]{ObergMurrayClayBergin2011, Molliere2020}. Since in DI the planets capture  a large, representative volume of the disk gas and dust simultaneously, the C/O ratio could remain close to the local disk value. However, accretion of solids with different volatile enrichment and from different locations in the disk (passing different "snowlines") could affect the final composition of the clumps. In addition, the local gas disk composition that provides the material for the planet can be quite different from the gas that makes up the host star. 

 \subsection{Do giant planets accrete mostly pebbles or planetesimals?}
A central debate in giant planet formation theory is whether the solid (heavy-element) mass is primarily supplied by the accretion of large, kilometer-scale planetesimals or the rapid capture of millimeter-to-centimeter-sized pebbles. Traditionally, these were viewed as competing models, but a modern consensus is emerging that formation is likely a multi-modal process. Observational evidence from ALMA and other high-resolution interferometers confirms that protoplanetary disks contain a continuous distribution of {\bf solids} sizes, ranging from fine dust and pebbles to larger bodies evidenced by gaps and rings \citep{Andrews2020,Birnstiel2023}.

Accounting for both populations is essential for a realistic model of core growth and envelope enrichment. Pebble accretion is highly efficient at high gas densities, allowing for the rapid assembly of a 
  core within the short lifetime of the gas disk—a vital requirement for forming giant planets around low-mass M-stars. However, as the planet grows and opens a gap, it can reach the pebble isolation mass and halt the flow of pebbles  \citep{LambrechtsJohansen2012}. In this regime, the accretion of planetesimals becomes the dominant mechanism for continued enrichment. These larger bodies are not as easily diverted by pressure gradients at the gap edge, allowing them to penetrate the envelope and contribute to the primordial composition gradients and "fuzzy core" structures. Thus, rather than an ``either/or" scenario, the formation of a giant planet is likely a sequential or concurrent integration of both solid reservoirs, where pebbles drive early rapid growth and planetesimals govern the final chemical composition and structure of gaseous planets. 
 

\subsection{Can we connect the measured  atmospheric composition with the bulk composition?}\label{sec2}
Connecting a giant planet's atmospheric composition to its bulk composition is one of the most significant, yet scientifically fraught, goals of modern planetary science. While traditionally treated as a direct proxy, several critical physical mechanisms suggest that the atmosphere may be a deceptive indicator of the planet's total inventory.

The fundamental assumption that atmospheric measurements (e.g., C/H or O/H ratios) represent the bulk composition requires the planet to be fully convective and fully-mixed. However, it is now clear that giant planets can be inhomogeneous in composition \citep{HelledStevenson2024, HelledHoward2024}. This inhomogeneity can be a result of the formation process \citep{VallettaHelled2020, Stevenson2022} or physical processes taking place during the planetary evolution such as phase separations and immisibillities. Consequently, the atmosphere only represents the final "veneer" of gas accreted during the disk's dissipation, potentially underestimating (or overestimating) the planet's total heavy-element mass by a significant amount.


We must also remember that current retrieval models from transit spectroscopy (e.g., JWST) are often limited to an altitude above $\sim$1 bar. This is a narrow "skin" compared to the thousands of kilometers of planetary radius. Relying on these shallow observations to infer the bulk metallicity ($Z_{planet}$) typically ignores the possibility of composition gradients and  fuzzy cores. Critically, if the primordial gradients described above persist, the atmosphere may only reflect the final "veneer" of gas accreted during the disk's dissipation. Alternatively, if a deep radiative layer separates the atmosphere from the bulk of the interior, the atmosphere can be more metal-rich due to the accretion of heavy elements during the planetary evolution   \citep{MuellerHelled2024_JupiterRadiative}. Furthermore, internal processes such as phase separation \citep{MankovichFortneyMoore2016,HowardMuellerHelled2024,HowardHelledBergermannRedmer2025}. For example, species like helium or neon are known to become insoluble in metallic hydrogen and "rain out" toward the deep interior can further deplete the observable upper atmosphere from certain elements. Of course the possibility of condensation and chemical interactions make the problem even more complex. 
Therefore, while atmospheric data provide a window into the formation environment (like snowline locations or C/O ratios),  relying solely on spectroscopy from facilities like JWST is insufficient for inferring the planetary bulk metallicity. 

\subsection{How do giant planets around M-stars form?}\label{sec2}
Although giant planets around low-mass stars  are rare, they do exist.
The formation of giant planets around small-mass stars, in particular, M-dwarf stars, represents a significant challenge to standard planet formation theory. This is because the protoplanetary disks around these stars are less massive, and the planetary growth rate is lower, which should theoretically inhibit the growth of gas giants. Under the CA model, the primary obstacle is the "growth-to-disk-lifetime" ratio; the low surface density of solids in M-dwarf disks means that building a 10 M$_{\oplus}$ 
  core often takes longer than the few million years the gas disk persists \citep{Burn2021}. However, the discovery of massive gas giants around low-mass stars, such as GJ 3512 b, has forced a re-evaluation of these limits. Even with pebble accretion which leads to a more efficient growth, planet formation models suggest that CA still struggles to produce Jupiters around stars with masses below 0.5 M$_{\odot}$
  unless the disks are exceptionally metal-rich \citep{ShibataHelled2025}. Of course,  given that these objects are rare, it is possible that they are the outcome of ``outlier" disks (e.g., massive, long-lived, metal-rich).

Alternatively, the DI  model is frequently invoked as the more likely pathway for these systems. Because DI occurs rapidly (on a scale of 1000  years) and depends on the disk's self-gravity rather than the slow accumulation of solids, it can bypass the time-scale constraints of core accretion \citep{BossKanodia2023}. In this scenario, the massive disk around a young M-star fragments directly into a gaseous clump. This is supported by the fact that giant planets around M-dwarfs seem to have lower bulk metallicities compared to giant planet orbiting sun-like stars \citep{MuellerHelled2025}, although this topic is still under investigation \citep{Chachanetal2025_Review}. There is also the possibility of a hybrid scenario, where a disk instability triggers a localized concentration of solids that then rapidly seeds a core. DI seems to be a rare but could be an effective ``shortcut" for creating the massive giant planets we observe around low-mass stars. 
 Although DI could  efficiently bypasses the formation time-scale constraint, it is probably  a rare occurrence, and the efficiency of DI around M-dwarfs remains unknown. 
 Overall, the exact formation pathways for giant planets orbiting low-mass stars remain elusive, as their existence frequently defies the predictive limits of standard formation theories. Consequently, these systems continue to challenge   existing theoretical frameworks, demanding a  reassessment of how gas giants emerge around small stars. 

\begin{figure}[h]
\centering
\includegraphics[width=0.9\textwidth]{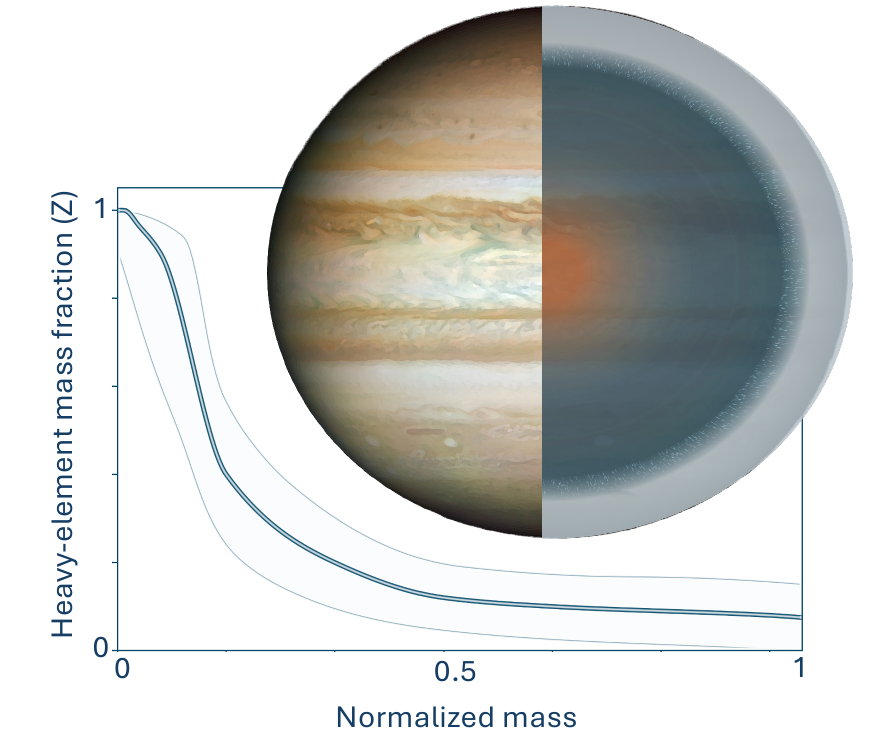}
\caption{A sketch of the planetary bulk metallicity (represented by the heavy-element mass fraction) as a function of planetary mass of a proto-Jupiter. The heavy-element mass fraction gradient is a result of the accretion history. It is therefore expected that low-mass planets are heavy-element dominated in composition, while planets more massive planets are H–He-rich. Also shown is the expected  internal structure of Jupiter with composition gradients (``fuzzy core") as predicted from formation and structure models.}  \label{fig1}
\end{figure}

\subsection{Can giant planets be differentiated in composition from birth?}\label{sec2}
Recent planet formation simulations clearly show that composition gradients are {\bf built up} during the planetary formation and that there is a clear link between the nature of the gradient and the accretion rates \citep{Helled2023, VallettaHelled2020}. At the same time, it seems  that ``primordial differentiation" of elements can also be produced.  This differentiation is caused  by the wildly different vaporization temperatures of different elements (i.e., metals, vs. rocks. vs. water). Also the conditions for fragmentation, which are linked to the material strength depend on the exact composition of the accreted solids. Therefore, differentiation in giant planets is fundamentally driven by the depth-dependent efficiency of thermal ablation and mechanical fragmentation as planetesimals traverse the forming envelope \citep{VallettaHelled2020,Stevenson2022}). Accreted  planetesimals do not deposit their mass uniformly; instead, volatile species with lower sublimation temperatures (such as water and methane) evaporate in the cooler, low-density outer layers, while refractory materials (silicates and metals) survive to significantly higher temperatures and pressures. This sorting is further reinforced by the physics of ram pressure ($P_{ram} \propto \rho v^2$), which increases with envelope density; because refractory-rich bodies typically possess higher material strength and density, they resist fragmentation longer than ice-rich planetesimals, penetrating deeper into the planetary interior before integrating their mass. Consequently, the planet is born not as a homogeneous mixture, but with a primordial composition gradient, where the gradient itself is  differentiated in composition, where the deep interior is inherently enriched in heavy, refractory elements, followed by an increase of more volatile materials \citep{VallettaHelled2021_CaptureRadius}. In other words, we expect that the composition gradients itself is differentiated in composition. It is yet to be determined whether this is indeed the case also when the accreted planetesimals are mixed in composition and are more similar to rubble piles.  Future research should determine whether such differentiation is not destroyed by large-scale convective mixing and whether the mixing timescale is comparable to the planetary age.  The efficiency of mixing, if operates, would depend on the exact  structure mean-molecular weight gradient, and the composition of the heavy elements \citep[e.g.,][and references therein]{Knierim2025}.

\subsection{How does convective mixing (and double diffusion) operate in giant planet interiors?}\label{sec2}

Convection in giant planets is the primary engine for both heat and material   transport and the generation of global magnetic fields via dynamo action. In a standard, adiabatic interior, convection operates as a large-scale, turbulent overturning of fluid driven by the planet's internal heat—remnant from formation.  Convective mixing however, is typically modeled using the in Mixing Length Theory (MLT), where the ``mixing length" is assumed to be proportional to the local pressure scale height.  Although MLT provides a useful (and rather elegant) framework for  estimating convective heat transport, the specific diffusion coefficient and mixing efficiency remain poorly constrained in the extreme high-pressure, high-temperature regimes of giant planet interiors. It is also possible that the MLT is not well suited for the conditions expected in giant planets. 

Another important issue is the modeling of convection with the presence of composition gradients. In that case, the efficiency of convection is drastically reduced, and the mixing length becomes a dynamic variable that depends on the strength of the stabilizing stratification.
When a stabilizing mean-molecular weight gradient (more heavy elements in the deeper interior) competes with a destabilizing thermal gradient (higher temperature in the deeper interior), large-scale convection may be suppressed in favor of double-diffusive convection, leading to the formation of convective staircases: a series of discrete, well-mixed layers separated by thin, stable interfaces. The efficiency of heat transport and mixing of elements under such conditions remains unknown.  It is also important to determine whether double-diffusive convection operates only transiently or over long time scales.
Finally, note that rotation and magnetic fields can also affect the efficiency of mixing and should not be ignored  \citep[e.g.,][]{FuentesAndersCummingHindman2023}. 
We hope that numerical simulations and lab experiments will provide more constraints on these parameters in the future.  




\section{Discussion \& Outlook}\label{sec13}
Giant planet science is currently undergoing a paradigm shift, moving from idealized, homogeneous interior models toward a framework that embraces complexity,  and more diverse and sophisticated formation and evolution  history. As of 2026, the synthesis of data from JWST, Juno \& Cassini as well as high-resolution ground-based surveys has provided a clearer, albeit more challenging, picture of these astrophysical objects. 
We are also at a stage where the debate between CA and DI is no longer viewed as a binary choice but as a continuum of possibilities: while core accretion remains the most robust explanation for the majority of observed giant planets, especially giant planets up to $\sim$ 1 M$_J$, disk instability has been elevated from a theoretical alternative to a demonstrated reality, and it is now considered the primary pathway for massive ``super-Jupiters" on wide orbits, and for giant planet formation around low-mass stars. 

Regarding giant planet interiors, a significant realization is that giant planets are inhomogeneous in composition, probably from birth. 
Since we now know that giant planets can be inhomogeneous in composition, we must be scientifically critical in the interpretation of "atmospheric metallicity." If composition gradients and boundary layers exist, the measured atmospheric composition may not represent the planetary bulk composition. 

An awareness of the current challenges in giant planet theory  that include non-adiabatic interiors, deceptive atmospheres, and hybrid formation paths, and several physical and chemical processes that can affect the final planetary composition and internal structure, will eventually allow us to map the history of our own Solar System and understand the diversity of giant planets observed around other stars. 
The open questions presented above represent only a few of those that should be addressed in the near future. It is clear that more topics need dedicated investigations, and also, that as we aim to solve questions new ones would arise. Nevertheless, these are exciting times as we no longer just discover planets, instead, we are beginning to read their biographies. 
\backmatter


\bmhead{Acknowledgements}
I thank my collaborators, students, and postdocs for interesting discussions. 

\section*{Declarations}
\begin{itemize}
\item Funding: Not Applicable 
\item Conflict of interest/Competing interests: Not Applicable
\item Ethics approval and consent to participate: Not Applicable
\item Consent for publication: Not Applicable
\item Data availability: Not Applicable
\item Materials availability: Not Applicable
\item Code availability: 
\item Author contribution: The author wrote the paper. 
\end{itemize}

\bibliography{sn-bibliography}

\end{document}